\documentclass[11pt]{article}
\usepackage{amsmath,amssymb,color,graphics,epsfig}
\usepackage{graphicx}
\usepackage{subfigure}
\usepackage{amsfonts}
\newcommand{\be}{\begin{equation}}
\newcommand{\ee}{\end{equation}}
\newcommand{\bea}{\setlength\arraycolsep{2pt} \begin{eqnarray}}
\newcommand{\eea}{\end{eqnarray}}
\newcommand{\nn}{\nonumber}

\def\ft#1#2{{\textstyle{\frac{\scriptstyle #1}{\scriptstyle #2} } }}
\def\fft#1#2{{\frac{#1}{#2}}}

\def\0{{\sst{(0)}}}
\def\1{{\sst{(1)}}}
\def\2{{\sst{(2)}}}
\def\3{{\sst{(3)}}}
\def\4{{\sst{(4)}}}
\def\5{{\sst{(5)}}}
\def\6{{\sst{(6)}}}
\def\7{{\sst{(7)}}}
\def\8{{\sst{(8)}}}
\def\sst#1{{\scriptscriptstyle #1}}

\begin{document}

\begin{flushright}
%\hfill{KIAS-P12028}
 %\hfill{
%\bf hep-th/yymmnnn}
\end{flushright}

\vspace{25pt}
\begin{center}
{\large {\bf On the Noether charge and the gravity duals of quantum complexity}}

\vspace{10pt}
 Zhong-Ying Fan$^{1\dagger}$, Minyong Guo$^{2,3\ast}$\\

\vspace{10pt}
$^{1\dagger}${ Center for Astrophysics, School of Physics and Electronic Engineering, \\
 Guangzhou University, Guangzhou 510006, China }\\
 $^{2\ast}${ Department of Physics, Beijing Normal University, \\
 Beijing 100875,  P. R. China}\\
$^{3\ast}${ Perimeter Institute for Theoretical Physics \\Waterloo, Ontario N2L 2Y5, Canada\\}
\smallskip
%{\it $^{2}$Department of Physics and State Key Laboratory of Nuclear Physics and Technology,\\}
%{\it Peking University, No.5 Yiheyuan Rd, Beijing 100871, P.R. China\\}
%\smallskip
%{\it $^{3}$Collaborative Innovation Center of Quantum Matter, No.5 Yiheyuan Rd,\\}
%{\it  Beijing 100871, P. R. China\\}

\vspace{40pt}

\underline{ABSTRACT}
\end{center}
The physical relevance of the thermodynamic volumes of AdS black holes to the gravity duals of quantum complexity was recently argued by Couch et al.
In this paper, by generalizing the Wald-Iyer formalism, we derive a geometric expression for the thermodynamic volume and relate its product with the thermodynamic pressure to the non-derivative part of the gravitational action evaluated on the Wheeler-DeWitt patch.
We propose that this action provides an alternative gravity dual of the quantum complexity of the boundary theory. We refer this to ``complexity=action 2.0" (CA-2) duality. It is significantly different from the original ``complexity=action" (CA) duality as well as the ``complexity=volume 2.0" (CV-2) duality proposed by Couch et al. The latter postulates that the complexity is dual to the spacetime volume of the Wheeler-DeWitt patch. To distinguish our new conjecture from the various dualities in literature, we study a number of black holes in Einstein-Maxwell-Dilation theories. We find that for all these black holes, the CA duality generally does not respect the Lloyd bound whereas the CV-2 duality always does. For the CA-2 duality, although in many cases it is consistent with the Lloyd bound, we also find a counter example for which it violates the bound as well.

\vfill {\footnotesize  Email: $^\dagger$fanzhy@gzhu.edu.cn\,,}
{\footnotesize   $^\ast$guominyong@gmail.com\,.}

\thispagestyle{empty}

\pagebreak

\tableofcontents
\addtocontents{toc}{\protect\setcounter{tocdepth}{2}}

%%%%%%%%%%%%%%%%%%%%%%%%%%%%%%%%%%%%%%%%

%\newpage
%%%%%%%%%%%%%%%%%%%%%%%%%%%%%%%%%%%%%%%%

%\vspace{2cm}

\section{Introduction}

Quantum computational complexity (or quantum complexity, in brief) is defined by the minimal number of quantum gates needed to build a target state of interest from a reference state. It is straightforward to see that any way to produce the state has already put constrains on the growth of the quantum complexity. In the field of quantum computations, it is believed that at exponentially late time, the growth of the complexity is linearly proportional to time. The proportional coefficient is postulated to be the double of the total energy of the system \cite{Lloyd}
\be \dot{\mathcal{C}}(|\psi(t)\rangle)\leq \fft{2E}{\pi \hbar} \,.\ee
This is known as the Lloyd bound. The linear growth at late time and the consistency with the Lloyd bound provide two guidelines for one to search the holographic duals of quantum complexity of the dual field theories.

\begin{figure}
  \centering
  	\subfigure{\includegraphics[width=3in]{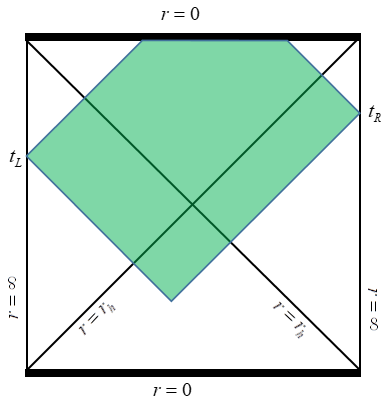}}
  	\caption{The Wheeler-DeWitt patch of a Schwarzschild black hole. The upper part of the patch runs into a singularity. }
\label{fig1}\end{figure}
The first holographic dual \cite{Stanford:2014jda} postulates that the complexity is dual to the volume of the maximal spacelike hypersurface crossing the Einstein-Rosen bridge. The proposal, known as ``complexity=volume" (CV) duality, correctly captures the linear growth of the complexity at late time but has a minor problem that a length scale should be properly chosen by hand in a case-by-case basis. This problem is cured by the ``complexity=action" or (CA) duality  \cite{Brown:2015bva,Brown:2015lvg}, which states that the complexity is dual to the gravitational action evaluated on the Wheeler-DeWitt (WDW) patch (the WDW patch of a Schwarzschild black hole is shown in fig.\ref{fig1})
\be \mathcal{C}=\fft{S}{\pi\hbar} \,,\ee
where the prefactor is chosen such that Schwarzschild-AdS black hole saturates the Lloyd bound. This is of course a conjecture as well but is fascinating since black holes are believed to be the fastest scramblers in nature \cite{Sekino:2008he}. The new proposal since then has attracted a lot of attentions in literature  (see, for example \cite{Lehner:2016vdi,Chapman:2016hwi,Carmi:2016wjl,Carmi:2017jqz,Chapman:2018dem,
Cai:2016xho,Kim:2017lrw,Kim:2017qrq,Yang:2016awy,Ruan:2017tkr,Moosa:2017yvt,Moosa:2017yiz,An:2018xhv}). However, it also has problems at the very start. As pointed out by Brown et al in \cite{Brown:2015bva,Brown:2015lvg}, for a Reissner-Nordstr\o m (RN) black hole the CA-duality generally does not respect the Lloyd bound\footnote{In fact, in this paper we investigate the CA-duality for a number of black holes in Einstein-Maxwell-Dilaton theories and find that they all do not respect the Lloyd bound}. In addition, to adopt the CA-duality one should deal with boundary terms, including joint terms appropriately. However, it is problematic for certain cases, such as magnetically charged black holes and higher derivative gravities (for recent developments, see \cite{Cano:2018ckq,Cano:2018aqi}).

Then is it possible that there exists an alternative and better holographic dual to the complexity? It was established in a very nice paper \cite{Couch:2016exn} that the spacetime volume of the WDW patch might be such an alternate choice. It was conjectured there that (dubbed by ``complexity=volume 2.0" or CV-2 duality)
\be \mathcal{C}=\fft{1}{\pi\hbar}\,P(\mathrm{Spacetime\, Volume}) \,,\ee
where $P$ denotes the thermodynamic pressure. This new proposal has several advantages: it is pure geometric but does not have the ad hoc of the original CV-duality; it is more easier for practical calculations because of no boundary terms needed; more importantly, in certain cases (for example, the RN black hole), it respects the Lloyd bound whereas the CA-duality does not.

In the same paper \cite{Couch:2016exn}, the authors also discuss the physical significance of the thermodynamic volumes of AdS black holes to the growth of the complexity. It was shown that in CA-duality the thermodynamic volume arises naturally in the calculations of the late time rate of change of the complexity. For a number of black holes in the Einstein-Maxwell theory (including pure Einstein gravity), they find that the thermodynamic volume simply equals to the late time growth of the spatial volume of the WDW patch but they have no proof for general cases.

Inspired by the discussions of \cite{Couch:2016exn}, we revisit how thermodynamic volume is derived in the generalized Wald-Iyer formalism. Our new finding is that generally the thermodynamic volume indeed has a geometric expression which is however not equivalent to the spatial volume of the WDW patch. In fact, its product with the thermodynamic pressure is nothing else but a part of the non-derivative action that is relevant to the cosmological constant (we call it $\Lambda$-action in brief). This makes it clear why the thermodynamic volume contributes to the growth of the complexity in the CA-duality. Considering its physical relevance to the holographic duals of quantum complexity, we propose that the $\Lambda$-action itself is dual to the complexity of the boundary field theories. To avoid confusions, we refer it to ``complexity=action 2.0" or CA-2 duality. To explore which one of the above dualities is the best holographic dual of the complexity, we study a number of black holes in Einstein-Maxwell-Dilaton (EMD) theories. We find that for all these black holes, the CA-duality generally does not respect the Lloyd bound whereas the CV-2 duality always does. For the CA-2 duality, though in many cases it is consistent with the Lloyd bound, we also find a counter example for which it violates the bound as well.

The paper is organized as follows. In section 2, we study the CA-duality for general static solutions in Einstein gravity. By using Wald-Iyer formalism, we prove that the late time growth of the action of the WDW patch can always split into two terms: one is the bulk action evaluated in the black hole interior and the other is the Gibbons-Hawking boundary term evaluated on the (inner and outer) event horizon. This greatly simplifies the calculations of complexity in CA-duality. In section 3, we derive a geometric expression for the thermodynamic volume by generalizing the Wald-Iyer formalism. Under some reasonable assumptions, we show that its product with the thermodynamic pressure essentially gives the non-derivative action that is linearly proportional to the cosmological constant. We thus propose the CA-2 duality. In section 4, we calculate the complexity by using the CA/CA-2/CV-2 dualities respectively for a variety of black holes in the EMD theories. We conclude in section 5.

\section{The Noether charge and CA duality}\label{sec2}

\subsection{The Noether charge and bulk action}
It was first developed by Wald and Iyer \cite{wald1,wald2} that for a generic gravity theory with diffeomorphism invariance, the thermodynamic first law of a stationary black hole can be systematically derived via the Noether charge associated to a time-like Killing vector field. The method is known as {\it Wald-Iyer formalism} (or {\it Wald formalism}) in literature. It provides a powerful tool to derive the first laws of black holes even if the solutions are numerical (see, for example \cite{Papadimitriou:2005ii,Lu:2013ura,Liu:2013gja,Liu:2014tra,Liu:2014dva,Fan:2014ixa,Lu:2014maa,Fan:2014ala,Fan:2015yza,Chen:2016qks,Feng:2015oea,
Feng:2015wvb,Fan:2016jnz,Fan:2017bka}). To begin our story, let us first review how the Noether charge is introduced in the {\it Wald-Iyer formalism}.

Variation of the action with respect to the dynamical fields, one finds
\be \delta \Big(\sqrt{-g}\mathcal{L} \Big)=\sqrt{-g}\Big(E_\Phi \delta \Phi+\nabla_\mu J^{\mu}\Big) \,,\ee
where $\Phi$ collectively denotes all the dynamical fields (the metric and matter fields) and $E_\Phi=0$ are the equations of motion. Without confusion, the tensor indices of $\Phi$ have been omitted for the sake of simplicity.
Note that the current $J^\mu$ depends linearly on the variation of the dynamical fields $\delta \Phi$. From this current $J^\mu$, one can introduce a current 1-form and a current (n-1)-form as follows
\be J_{(1)}:=J_\mu dx^\mu,\qquad \Theta_{(n-1)}:=*J_{(1)} \label{joneform}\,.\ee
The Noether current $(n-1)$-form is defined by:
\be J_{(n-1)}:=\Theta_{(n-1)}-\xi_{(1)}\cdot *\mathcal{L} \,,\label{noetherj}\ee
where $\xi_{(1)}\cdot$ denotes the contraction of $\xi$ to the first index of the tensor it acted upon. It was first shown by Wald \cite{wald1} that
\be dJ_{(n-1)}=-E_{(n)}\delta \Phi= 0 \label{dnoetherj}\,,\ee
where $E_{(n)}$ denotes the $n$-form equations of motion. Thus, one can define a Noether charge (n-2)-form as
\be J_{(n-1)}:= dQ_{(n-2)}+E.O.M \label{noethercharge}\,,\ee
where $E.O.M$ denotes the terms proportional to the equations of motion. Then taking the dual form of the equation (\ref{noetherj}), one finds
\be *d*\mathcal{Q}_{(2)}=*J_{(n-1)}=*\Theta_{(n-1)}-*\Big(\xi_{(1)}\cdot *\mathcal{L} \Big) \,,\label{central1}\ee
where $\mathcal{Q}_{(2)}$ is the Noether charge 2-form and we have dropped the terms of the equations of motion. This gives rise to
\be \nabla_\nu \mathcal{Q}^{\mu\nu}=J^\mu-\xi^\mu\mathcal{L} \,.\label{central2}\ee
Here it is worth emphasizing that the above Noether charge is defined for any vector field $\xi$ that is not limited to a Killing vector field. However, when $\xi$ is a Killing vector field, one has $\delta \Phi=L_\xi \Phi=0$ leading to $J^\mu(\delta\Phi)=0$ since the current $J^\mu$ depends linearly on $\delta \Phi$. With this observation, it was shown in \cite{Liu:2017kml,Li:2017ncu,Fan:2018qnt} that the Noether charge of a time-like/space-like Killing vector field is dual to the heat current/shear stress in the boundary. This greatly simplifies the derivation of thermal-electric conductivities as well as the ratio of shear viscosity to the entropy density in the AdS/CFT correspondence.

For later purpose, we consider generally static solutions with spherical/hyperbolic/toric isometries. By taking $\xi=\partial/\partial t$, we find
\be\label{bulk} \partial_r\Big(\sqrt{-g}\mathcal{Q}^{rt} \Big)=\sqrt{-g}\,\mathcal{L} \,,\ee
 where the r.h.s of the equation is nothing else but the Lagrangian of the on-shell solutions. Hence, the bulk action defined on any space-time region can split into several boundary terms of the same region. This plays an important role in our calculations of the action growth for generally static solutions, as will be shown later.

\subsection{The action growth and CA-duality}

In this subsection, we will calculate the growth of quantum complexity of the dual field theories by using the CA-duality. The CA-duality relates the complexity of the boundary CFT state to the gravitational action evaluated on the Wheeler-DeWitt (WDW) patch (see Fig.\ref{fig1}). For simplicity, we will drop the overall factor $1/\pi\hbar$ and focus on the growth of the action at late time in the following. In literature, it is known that the computation of the action growth of the WDW patch can be performed by using two different methods which were developed in \cite{Brown:2015bva,Brown:2015lvg} and \cite{Lehner:2016vdi} respectively. In particular, the latter is more rigorous with a more careful treatment of the boundary terms of the WDW patch. It has been found that the two methods give the same results for the Schwarzschild and Reissner-Nordstr\o m black holes. Here our new contribution is we will prove the two methods are in fact always equivalent for generally static solutions with maximal symmetries, even if without knowing the details of the solutions.

 To be concrete, we focus on generally electrically charged static solutions in Einstein-Maxwell-Dilaton (EMD) theories. The bulk action is given by
\be
S_{bulk}=\fft{1}{16\pi G}\int_{\mathcal{M}} \mathrm{d}^{n}x\sqrt{-g}\Big(R-\ft {1}{4e^2} Z(\phi)F^2-\ft 12 \big(\partial \phi\big)^2-U(\phi)  \Big)\,.
\ee
We take a general ansatz for the static solution
\be\label{static} ds^2=-h(r) dt^2+dr^2/f(r)+\rho(r)^2 dx^i dx^i\,,\quad A=a(r) dt\,,\quad \phi=\phi(r) \,.\ee

\subsubsection{Without joints}

We shall first calculate the action growth using the method in \cite{Brown:2015bva,Brown:2015lvg}. The total action is given by the sum of the bulk action and the Gibbons-Hawking (GH) boundary term
\be\label{onshellaction} S=S_{bulk}+\fft{1}{8\pi G}\int_{\partial \mathcal{M}}\mathrm{d}^{n-1}y\sqrt{-h}K \,.\ee
 The key ingredient of the calculations of \cite{Brown:2015bva,Brown:2015lvg} is that the late time growth of the action of the WDW patch is essentially given by the bulk action evaluated in the black hole interior, together with the GH boundary term evaluated on the (inner and outer) event horizon. The null boundaries and joints do not have effective contributions. Hence, integrating the equation (\ref{bulk}) on the WDW patch, one finds
\be\label{bulk2} \dot S_{bulk}\equiv \fft{dS_{bulk}}{dt_L}=\fft{\omega_{n-2}}{16\pi G}\int_{r_-}^{r_+}\mathrm{d}r \sqrt{-\bar g}\,\mathcal{L}=\fft{\omega_{n-2}}{16\pi G}\sqrt{-\bar g}\,\mathcal{Q}^{rt}\Big|^{r_+}_{r_-}\,, \ee
where $\omega_{n-2}\equiv \int_{\Sigma_{n-2}}\sqrt{\gamma}\, dx^1 dx^2\cdots dx^{n-2}$ denotes the volume factor of the codimension-2 subspace and $\bar g\equiv g/\gamma$. $r_\pm$ denote the outer and inner horizon of the black hole ($r_-\rightarrow 0$ when the inner horizon is absent or in other words, the singularity is viewed as the inner horizon of the black hole). It should be emphasized that the above result is formal and valid to general gravitational theories with diffeomorphism invariance. For the EMD theories, the Noether charge 2-form is given by
\be \mathcal{Q}^{\mu\nu}=-2\nabla^{[\mu}\xi^{\nu]}-\ft{Z(\phi)}{e^2}F^{\mu\nu}A_\sigma \xi^\sigma \,.\ee
Then substituting the static solution (\ref{static}) into the action, we deduce
\bea
&&\dot S_{bulk}=-\fft{\omega_{n-2}}{16\pi G}\,\rho^{n-2}\sqrt{hf}\Big( \ft{h'}{h}-\ft{Z}{e^2}\ft{a a'}{h} \Big)\Big|^{r_+}_{r_-}\,,\nn\\
&&\dot S_{GH}=\fft{\omega_{n-2}}{16\pi G}\,\rho^{n-2}\sqrt{hf}\Big( \ft{h'}{h}+\ft{2(n-2)\rho'}{\rho} \Big)\Big|^{r_+}_{r_-}\,.
\eea
Thus, the growth of the total action is
\be\label{sdot} \dot S=\fft{\omega_{n-2}}{16\pi G}\,\rho^{n-2}\sqrt{hf}\Big( \ft{Z}{e^2}\ft{a a'}{h}+\ft{2(n-2)\rho'}{\rho} \Big)\Big|^{r_+}_{r_-}\,.   \ee

To verify the above result, we first consider a Schwarzschild black hole which has
\be  a=0=\phi\,,\quad \rho=r\,,\quad  h=f=r^2\ell^{-2}+k-\fft{16\pi}{(n-2)\omega_{n-2}}\fft{G M}{r^{n-3}}\,.\ee
We easily find $\dot S=2M$, which saturates the Lloyd bound. For a Reissner-Nordstr\o m (RN) black hole
\bea
&&\phi=0\,,\quad \rho=r\,,\quad a= \mu-\fft{16\pi e^2}{(n-3)\omega_{n-2}}\fft{G Q}{r^{n-3}}\,,\nn\\ &&h=f=r^2\ell^{-2}+k-\fft{16\pi}{(n-2)\omega_{n-2}}\fft{G M}{r^{n-3}}+\fft{128\pi^2 e^2}{(n-2)(n-3)\omega_{n-2}^2}\fft{G^2 Q^2}{r^{2(n-3)}}\,,
\eea
we obtain
\be \dot S=\fft{16\pi G e^2}{(n-3)\omega_{n-2}}\Big(\fft{Q^2}{r_-^{n-3}}-\fft{Q^2}{r_+^{n-3}} \Big) \,.\ee
These results are exactly matched with those published in \cite{Brown:2015bva,Brown:2015lvg}.
\begin{figure}
  \centering
  	\subfigure{\includegraphics[width=3in]{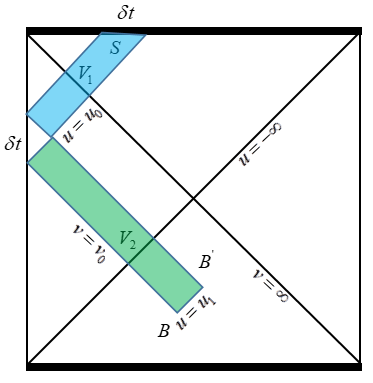}}
  	\caption{The action growth of the WDW patch of an uncharged black hole. In the time process $t_L\rightarrow t_L+\delta t$, the patch loses a sliver $V_2$ (Blue) and gains another one $V_1$ (Green). }\label{wj}
 \end{figure}

\subsubsection{With joints}

Despite that the method in \cite{Brown:2015bva,Brown:2015lvg} looks simple, the contributions of the joints at which the null boundaries intersect are totally ignored. This was questioned in \cite{Lehner:2016vdi}. In fact, to compute the action growth of the WDW patch more rigorously, one should treat the joint terms more carefully. In an infinitesimal time process $t_L\rightarrow t_L+\delta t$, the action growth of the WDW patch $\delta S=S(t_L+\delta t)-S(t_L)$ takes the form of
\be\label{rm}
\delta S=S_{V_1}-S_{V_2}-\frac{1}{8\pi G}\oint_{S}Kd\Sigma+\frac{1}{8\pi G}\oint_{B^{\prime}}bdS-\frac{1}{8\pi G}\oint_{B}bdS\,,
\ee
where $V_1$ and $V_2$ denote the upper and lower slivers in Fig.\ref{wj}, respectively. $\oint_S K d\Sigma$ is the GH boundary term evaluated at the black hole singularity and $b=\log\lvert k\cdot \bar{k}\rvert$, where $k$ and $\bar{k}$ are the two null normals to the corner pieces. For the following calculations, it will be convenient to introduce the null coordinates $u$ and $v$ defined by
\bea
u&=&t+r_*\,,\nn\\
v&=&t-r_*\,,
\eea
where $r_*(r)=\int dr(hf)^{-1/2}$ is the tortoise coordinate. Thus, the metric can be written as
\bea
ds^2=-hdu^2+2(h/f)^{1/2}dudr+\rho(r)^2d\Omega_{n-2\,,k}^2\,,
\eea
in the ingoing coordinate and
\bea
ds^2=-hdv^2-2(h/f)^{1/2}dvdr+\rho(r)^2d\Omega_{n-2\,,k}^2\,,
\eea
in the outgoing coordinate, respectively. In addition, note that $dt\wedge dr=du\wedge dr=dv\wedge dr$, we have a simple and useful relation
\be
\int d^nx\,\sqrt{-g}=\int d\Omega_{n-2}\,dw dr\,\sqrt{-g}=\omega_{n-2}\int dw dr\,\sqrt{-\bar g}\,,
\ee
where $w$ collectively denotes the various time coordinates $t,u,v$.

First, we consider the difference of the action between the two rectangles $S_{V_1}-S_{V_2}$. We have
\bea
S_{V_1}&=&\frac{1}{16\pi G}\int_{u_0}^{u_0+\delta t}du\int d\Omega_{n-2} \int^{\lambda(u)}_{\epsilon}dr\sqrt{-g}\mathcal{L}\,,\nn\\
&=&\frac{\omega_{n-2}}{16\pi G}\int_{u_0}^{u_0+\delta t}du\, \big[\sqrt{-\bar g}\mathcal{Q}^{rt}\big]_{\epsilon}^{\lambda(u)}\,,
\eea
where in the first \lq\lq{}=\rq\rq{}, we denote $r=\lambda(u)$ to describe the null hypersurface $v=v_0+\delta t$. The function $\lambda(u)$ can be solved by the equation $r_*\big(\lambda(u)\big)=\frac{1}{2}(u-v_0-\delta t)$. In the second \lq\lq{}=\rq\rq{}, we have used Eq.(\ref{bulk}). Similarly, for the rectangle $V_2$, we have
\bea
S_{V_2}&=&\frac{1}{16\pi G}\int_{v_0}^{v_0+\delta t}dv\int d\Omega_{n-2}\int^{\lambda_0(v)}_{\lambda_1(v)}dr\sqrt{-g}\mathcal{L}\,,\nn\\
&=&\frac{\omega_{n-2}}{16\pi G}\int_{v_0}^{v_0+\delta t}dv\, \big[\sqrt{-\bar g}\mathcal{Q}^{rt}\big]_{\lambda_1(v)}^{\lambda_0(v)}\,.
\eea
Here we define $r=\lambda_{0,1}(v)$ to describe the null hypersurfaces $u=u_{0\,,1}$. They are determined by $r_*\big(\lambda_{0\,,1}(v) \big)=\frac{1}{2}(u_{0\,,1}-v)$.
Then changing the variables $u=u_0+v_0+\delta t-v$ in the integral of $S_{V_1}$, we find
\bea
S_{V_1}&=&\frac{\omega_{n-2}}{16\pi G}\int_{v_0}^{v_0+\delta t}dv \big[\sqrt{-\bar g}\mathcal{Q}^{rt}\big]_{\epsilon}^{\lambda_0(v)}\,,
\eea
where we have identified $\lambda_0(v)=\lambda(u)$ since they both describe a same radii at which the null boundaries $u=u_0$ and $v=v_0+\delta t$ intersect.
Combining the above results, we deduce
\bea
S_{V_1}-S_{V_2}&=&\frac{\omega_{n-2}}{16\pi G}\int_{v_0}^{v_0+\delta t}dv \,\big[\sqrt{-\bar g}\mathcal{Q}^{rt}\big]_{\epsilon}^{\lambda_1(v)}\,.
\eea
Considering the radii $r=\lambda_1(v)$ varies from $r_B$ to $r_{B^{\prime}}$ as $v$ increases from $v_0$ to $v_0+\delta t$, we find $r_{B^{\prime}}=r_{B}+O(\delta t)$ since the variation of the radius is very small. In particular, at late time, $r_{B}$ approaches to $r_+$. So the volume contribution to $\delta S$ is
\be\label{bt}
S_{V_1}-S_{V_2}=\frac{\omega_{n-2}}{16\pi G}\left[\sqrt{-\bar g}\mathcal{Q}_{tot}^{rt}\right]_{\epsilon}^{r_+}\delta t=\dot{S}_{bulk}\delta t\,.
\ee
This is consistent with Eq.(\ref{bulk2}). Then comparing Eq.(\ref{rm}) with Eq.(\ref{onshellaction}), we still need prove the joint terms $\pm2\oint bdS$ contributed by the codimension-2 surfaces $B$ and $B^{\prime}$ reproduce the GH boundary term evaluated at the event horizon.

Following the conventions in \cite{Lehner:2016vdi}, we write the null normals as
\bea
k_\alpha&=&-c\partial_\alpha v=-c\partial_\alpha(t-r^{\ast})\nn\\
\bar{k}_\alpha&=&\bar{c}\partial_\alpha u=\bar c\partial_\alpha (t+r^{\ast})
\eea
where $c$ and $\bar{c}$ are arbitary positive  constants which can be fixed by implementing the asymptotic normalizations $k\cdot \hat{t}_L=-c$ and $\bar{k}\cdot\hat{t}_R=-\bar{c}$, where $\hat{t}_{L,R}$ are the asymptotic Killing vectors which are normalized to describe the time flow in the left and right boundary theories, respectively. With these choices, we have $k\cdot\bar{k}=-2c\bar{c}/h$, so that
\bea
b=-\log\Big(\frac{h}{2c\bar{c}}\Big)\,.
\eea
We deduce
\bea
\frac{1}{8\pi G}\Big(\oint_{B^{\prime}}bdS-\oint_{B}bdS\Big)=\frac{\omega_{n-2}}{8\pi G}\Big( p(r_{B^{\prime}})-p(r_{B}) \Big)\,,
\eea
where $p(r)=-\rho^{n-2} \log\Big( \fft{h}{2c\bar c}\Big)$. Since the variation between $r_{B}$ and $r_{B^{\prime}}$ is very small, we can perform a Taylor expansion for $p(r)$ around $r=r_B$. Note that the displacement is in the direction of the $v$-axis. We have $du=0$, $dv=\delta t$, and $dr=-\frac{1}{2}\sqrt{hf}\delta t$. This gives rise to
\bea
p(r_{B^{\prime}})-p(r_{B})&=&\frac{\delta t}{2}\Big((n-2)\sqrt{hf}\rho^{n-3}\rho^{\prime}\log\big(\ft{h}{2c\bar c}\big)+\sqrt{\ft{f}{h}}\rho^{n-2}h^{\prime}\Big)_{r=r_B}\,.
\eea
Evaluating this at late time ($r_B$ approaches to $r_+$, $f$ and $h$ approach to zero), we obtain
\bea\label{jt}
\frac{1}{8\pi G}\Big(\oint_{B^{\prime}}bdS-\oint_{B}bdS \Big)&=&\frac{\omega_{n-2}}{16\pi G}\Big(\sqrt{hf}\rho^{n-2}\big( \log h \big)^{\prime}\Big)_{r=r_+}\nn\\
&=&\frac{1}{8\pi G}\int_{+}Kd\Sigma\,,
\eea
exactly as expected. This completes our proof for the uncharged black holes.

\begin{figure}
  \centering
  	\subfigure{\includegraphics[width=2in]{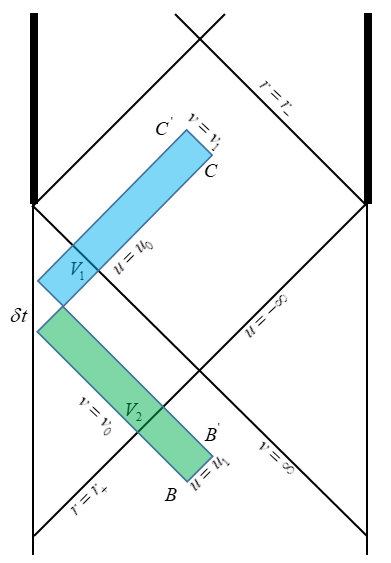}}
  	\caption{The WDW patch of a charged AdS black hole with an inner horizon. The upper part of the patch runs into the inner horizon at late time, rather than a singularity. Thus, there are some new joints $C\,,C'$ which also contribute to the action growth.}\label{wc}
 \end{figure}

\subsubsection{Extension to charged black holes}
 We can extend our previous results to charged AdS black holes with an inner horizon. The WDW patch is shown in Fig. \ref{wc}. It is easily seen that there are some new joints $C$ and $C^{\prime}$ contributing to the action growth. We have
\bea
\delta S=S_{V_1}-S_{V_2}+\frac{1}{8\pi G}\oint_{B^{\prime}}bdS-\frac{1}{8\pi G}\oint_{B}bdS+\frac{1}{8\pi G}\oint_{C^{\prime}}b^{\prime}dS-\frac{1}{8\pi G}\oint_{C}b^{\prime}dS\,,
\eea
where $b^{\prime}=-b$. As previously, the volume contribution to the action at late time is given by the bulk action evaluated at the black hole interior. We have
\bea
\delta S_V&=&\frac{\omega_{n-2}}{16\pi G}\int_{u_0}^{u_0+\delta t}du \big[\sqrt{-\bar g}\mathcal{Q}_{tot}^{rt}\big]_{\lambda_1(v)}\nn\\
&&-\frac{\omega_{n-2}}{16\pi G}\int_{v_0}^{v_0+\delta t}dv\big[\sqrt{-\bar g}\mathcal{Q}_{tot}^{rt}\big]_{\lambda^1(u)}\nn\\
&=&\frac{\omega_{n-2}}{16\pi G}\big[\sqrt{-\bar g}\mathcal{Q}_{tot}^{rt}\big]_{r_-}^{r_+}=\dot{S}_{bulk}\delta t\,,
\eea
where we define $r=\lambda_{1}(v)$ and $r=\lambda^{1}(u)$ to describe the null surfaces $v=v_1$ and $u=u_1$ respectively. Likewise, according to our previous discussions, the joint contributions read
\bea
\delta S_{B, B^{\prime}}&=&\frac{\omega_{n-2}}{16\pi G}\Big(\sqrt{hf}\rho^{n-2}\big(\log h \big)^{\prime}\Big)_{r=r_+}\,,\nn\\
\delta S_{C, C^{\prime}}&=&-\frac{\omega_{n-2}}{16\pi G}\Big(\sqrt{hf}\rho^{n-2}\big( \log h \big)^{\prime}\Big)_{r=r_-}\,.
\eea
Thus,
\bea
\delta S_{B, B^{\prime}}+\delta S_{C, C^{\prime}}=\frac{1}{8\pi G}\oint_{\pm}Kd\Sigma\,.
\eea
We again reproduce Eq.(\ref{onshellaction}), although the joint contributions are treated more carefully.

However, mathematically this is a more rigorous method to calculate the growth of the complexity of the dual field theories. In particular, the method itself is not limited to the late time. For example, it can be applied to the calculations of the full time evolution of the complexity in a dynamical space-time whereas the method of \cite{Brown:2015bva,Brown:2015lvg} cannot (we refer the readers to \cite{Chapman:2018dem} for details on this topic).

Furthermore, though we focus on Einstein gravity in this paper, the above discussions show some universal features so that it may be generalized to higher derivative gravities. The total action growth of the WDW patch is given by two pieces of contributions: one is the bulk action of the black hole interior and the other is the joint terms
\be \delta S_{WDW}=\delta S_{bulk}\big|_{r_-}^{r_+}+\delta S_{joint} \,.\ee
 For Einstein gravity, we have proved that at late time $\delta S_{joint}\rightarrow \big(\delta S_{GH}\big|_{r_+}-\delta S_{GH}\big|_{r_-}\big)$. However, for general higher derivative gravities, it is of great difficult to derive the specific form of the joint terms. Nonetheless, we notice that the late time limit is in some sense equivalent to the smooth limit of the boundaries. If this is correct, the above argument should not rely on the details of the gravity theories as well as the form of the solutions.

 To discuss this, we first recall that the (smooth) boundary action $S_{BDY}$ for a gravitational theory is derived by the variational principle. Likewise, when the boundary is non-smooth, the contributions of the joints should be included if the variational principle is well defined. Thus,
\be S_{BDY}\simeq \sum_{\partial\mathcal{M}_i}S^{(i)}_{BDY}+S_{joints} \,,\ee
where ``$\simeq$" means the equation is valid in the smooth limit. When the bulk region is the WDW patch, $\partial \mathcal{M}_i$ are the corresponding null boundaries which however do not contribute to the action growth. So we may conclude
\be \delta S_{BDY}\simeq \delta S_{joints}  \,,\ee
at the late time limit. This does not depend on any details of the theories and the solutions.

Thus, we argue that for general higher derivative gravities, the late time growth of the action of the WDW patch is simply given by
\be \delta S_{WDW}=\delta S_{bulk}\big|_{r_-}^{r_+}+\delta S_{BDY}\big|_{r_+}-\delta S_{BDY}\big|_{r_-}\,.\ee

\subsection{Application to hairy black holes}\label{application}

 Now we would like to compute the action growth for a number of scalar hairy black holes using Eq.(\ref{sdot}). The results will be compared with the growth of the complexity derived from different proposals (see sec.\ref{dualities}). The Lagrangian density of Einstein-Scalar gravity takes the form of
\be \mathcal{L}=R-\ft 12\big(\partial \phi \big)^2-U(\phi) \,,\ee
where $U(\phi)$ is the scalar potential which has a small $\phi$ expansion as $U=2\Lambda+\ft 12 m^2 \phi^2+\cdots$. For later convenience, we parameterize the cosmological constant as $\Lambda=-\ft 12 (n-1)(n-2)g^2$, where $g=1/\ell$ is the inverse of AdS radius. We find that for scalar hairy black holes, the action growth of the WDW patch at late time takes the form of
\be \dot S=2\delta\, M \,,\ee
where $\delta$ is no longer an universal constant. It depends on the parameters of theories as well as those of the solutions.

For example, for the scalar hairy black hole found in \cite{Gonzalez:2013aca}, one has
\bea\label{zxf}
&&U=-2g^2\big(\cosh{\phi}+2 \big)-2\alpha^2 \Big( \phi\big(\cosh{\phi}+2 \big)-3\sinh{\phi}  \Big)\,,\nn\\
&&ds^2=-f dt^2+dr^2/f+r(r+q)d\Omega_{2\,,k}^2\,,\quad \phi=\log{\Big(1+\fft{q}{r} \Big)}\,,\nn\\
&&f=g^2r^2+(g^2-\alpha^2)q r+k-\ft 12 \alpha^2 q^2+\alpha^2\, r(r+q)\log{\Big(1+\fft{q}{r} \Big)}\,.
\eea
By simple calculations, we obtain
\be \dot S=3M\Big(1-\ft k2 \big(\ft{\omega_{n-2}}{6\pi\alpha G M} \big)^{2/3}\Big) \,.\ee

For a different potential
\be
U(\phi)=-(g^2-\alpha^2) V_0(\phi) - \alpha^2 V_0(-\phi)\,,
\ee
with
\be
V_0(\phi) = (1+\mu)(1+2\mu) e^{-\sqrt{\fft{1-\mu}{1+\mu}}\,\phi} +
(1-\mu)(1-2\mu)  e^{\sqrt{\fft{1+\mu}{1-\mu}}\,\phi} + 4(1-\mu^2) e^{\fft{\mu}{\sqrt{1-\mu^2}}\,\phi}\,.
\ee
there exists a class of hairy black holes \cite{Fan:2015oca}
 \bea\label{zxf2}
&&ds^2=-f dt^2+dr^2/f+r^{\fft{1+\mu}{2}}(r+q)^{\fft{1-\mu}{2}}d\Omega_{2\,,k}^2\,,\quad \phi=\sqrt{1-\mu^2}\,\log{\Big(1+\fft{q}{r} \Big)}\,,\nn\\
&&f=\big(g^2-\alpha^2\big)\Big(1+\fft q r\Big)^\mu \Big(r^2+(1-2\mu)q r-\mu(1-2\mu)q^2 \Big)\nn\\
&&\qquad+k\Big(1+\fft q r \Big)^\mu+\alpha^2 r^{1+\mu}(r+q)^{1-\mu}\,,
\eea
where the parameter $0\leq\mu<1$. The growth of the action is given by
\be \dot S= \ft{3(1+\mu)}{1+2\mu}M\Big(1-\ft{(1-\mu)k q\,\omega_{n-2}}{24\pi G M} \Big)  \,.\ee
From these two examples, one finds that the action growth of spherical/planar hairy black holes does not respect the Lloyd bound\footnote{We also note that the action growth of the hyperbolic black holes violates any upper bound of the form $\dot S\leq \beta M$. In many cases, this happens as well for the CA-2 and CV-2 dualities. These results may suggest that the WDW patch of hyperbolic black holes is not dual to the complexity of the boundary theories. So we shall focus on spherical/planar black holes in this paper.}. They may satisfy a modified upper bound $\dot S\leq 3M$. However, this is not true. We consider the hairy planar black holes in \cite{Fan:2015ykb}. The scalar potential is given by
\bea U&=&-\ft 12 \big(\cosh{\Phi}\big)^{\ft{\mu k_0^2}{n-2}}\Big(g^2-\alpha^2 \big(\sinh{\Phi}\big)^{\ft{n-1}{\mu}} {}_2F_1\big(\ft{n-1}{2\mu},\ft{\mu k_0^2}{4(n-2)},\ft{n+2\mu-1}{2\mu},-\sinh^2{\Phi}  \big)   \Big)\\
&\times& \Big(2(n-2)(n-1)-\mu^2k_0^2\tanh^2{\Phi} \Big)-\alpha^2 (n-2)(n-1)\big(\cosh{\Phi}\big)^{\ft{\mu k_0^2}{2(n-2)}}\big(\sinh{\Phi}\big)^{\ft{n-1}{\mu}}\,,\nn
\eea
and the solutions read
\bea\label{fan}
&&ds^2=-fdt^2+\fft{\sigma^2dr^2}{f}+r^2dx^idx^i\,,\qquad\quad \phi=k_0\,\mathrm{arcsinh}(\ft{q^\mu}{r^\mu})\,,\nn\\
      &&\sigma=\Big(1+\ft{q^{2\mu}}{r^{2\mu}} \Big)^{-\ft{\mu k_0^2}{4(n-2)}}\,,\qquad f=g^2 r^2-\ft{\alpha^2\, q^{n-1}}{r^{n-3}}
      {}_2F_1\Big(\ft{n-1}{2\mu},\ft{\mu k_0^2}{4(n-2)},\ft{n+2\mu-1}{2\mu},-\ft{q^{2\mu}}{r^{2\mu}} \Big)   \,,
\label{staticsol}\eea
We deduce
\be \dot S=2\delta M\,,\quad \delta=\ft{2(n-1)(n-2)}{2(n-1)(n-2)-\mu^2 k_0^2}>1\,,\ee
where $\mu^2 k_0^2<2(n-1)(n-2)$ is a sufficient condition to guarantee the existence of the event horizon  \cite{Fan:2015ykb}. In this case, by tuning the parameters appropriately, one can have an arbitrary large value for $\delta$ and hence $\dot S$ will violate any modified upper bound.

\section{Thermodynamic volume and complexity}

\subsection{Thermodynamic volume from Wald-Iyer formalism}

It was shown in \cite{Couch:2016exn} that by generalizing the Wald-Iyer formalism, the thermodynamic volume can be calculated for any stationary black hole solutions. In fact, the discussions there are valid for all the coupling constants (denoted by $\lambda_\alpha$) in the Lagrangian. In the extended phase space, the variation of the action reads
\be \delta \Big(\sqrt{-g}\mathcal{L} \Big)=\sqrt{-g}\Big(E_\Phi \delta \Phi+\nabla_\mu J^{\mu}+\fft{\partial \mathcal{L}}{\partial \lambda_\alpha}\delta \lambda_\alpha\Big) \,,\ee
where the summation over $\alpha$ index is understood. Following the standard procedure, it is easy to show that
\be\label{first} \delta H_\infty=\delta H_+-\fft{1}{16\pi G}\int_\Sigma \xi_{(1)}\cdot \fft{\partial \mathcal{*L}}{\partial \lambda_\alpha}\delta \lambda_\alpha \,,\ee
where
\be \delta H=\frac{1}{16\pi G}\int_{\mathcal{A}_{n-2}}\Big( \delta Q_{\sst{(n-2)}}-\xi_{(1)}\cdot \Theta_{\sst{(n-1)}} \Big) \,,\ee
and $\Sigma$ is a Cauchy surface with two boundaries $\mathcal{A}_{n-2}$, one on the horizon and the other at infinity. Without the variation of the coupling constants $\lambda_\alpha$, the above equation gives the original thermodynamic first laws of stationary black holes. Hence, roughly speaking, the equation (\ref{first}) generalizes the first laws of black holes in the extended phase space so that one may systematically introduce new pairs of thermodynamic conjugates. However, there still exists some subtleties. For example,
if we take $\lambda_\alpha$ to be the cosmological constant $\Lambda$, the integral on the r.h.s of Eq.(\ref{first}) will diverge at the asymptotic infinity. This should be treated carefully. In fact, in view of the first law of thermodynamics, a same divergent term should emerge in $\delta H_\infty$ as well so that the final result of Eq.(\ref{first}) is still finite. This can be checked in a case-by-case basis, as shown in \cite{Couch:2016exn}. Thus, we propose
\bea\label{waldgene}
&&\delta H_\infty^{reg}=\delta H_++\widetilde V \delta P \,,\nn\\
&&P=-\fft{\Lambda}{8\pi G}\,,\quad \widetilde V=-\fft 12\int^{r_+} \xi_{(1)}\cdot \fft{\partial \mathcal{*L}}{\partial \Lambda}\,,
\eea
where $\delta H_\infty^{reg}=\delta H_\infty-\mathrm{divergent\,\,terms}$ and $\int^{r_+}$ is understood as the integral is regularized by removing the divergent terms at the asymptotic infinity. It should be emphasized that the above volume $\widetilde V$ is in general not identical to the thermodynamic volume defined in the usual way $V=\big( \partial M/\partial P \big)_{S}$. The two quantities may differ by some constant terms independent of $r_+$ (before the horizon condition $h(r_+)=0$ substituted into the integral). We introduce an illusive radius $r_i$ such that
\be\label{volume0}  V=-\fft 12\int^{r_+}_{r_i} \xi_{(1)}\cdot \fft{\partial \mathcal{*L}}{\partial \Lambda} \,.\ee
Of course, the value of the radii $r_i$ should be solved from this equation. We find that it could be either zero or non-zero and even has multivalues. However, these extra degrees of freedom can be removed by imposing proper ``boundary conditions". For example, when the black hole does not have an inner horizon, we require $r_i=0$ because in this case the thermodynamic volume defined at the singularity should vanish (recall that in this case, the singularity can be taken as the inner horizon of the black hole). Indeed, we can verify this very carefully for all the black holes studied in this paper.
On the other hand, when the black hole has an inner horizon, one can define the thermodynamic volume $V_-$ by replacing $r_+\rightarrow r_-$ in Eq.(\ref{volume0}) without changing the value of $r_i$. However, as will be shown later, in this case the detail of $r_i$ is irrelevant for our discussions.

 From Eq.(\ref{volume0}), we will show that the thermodynamic volume is closely related to the non-derivative part of the gravitational action evaluated on the WDW patch under some reasonable assumptions.

%\footnote{There is a minor mistake in the calculations [xxx] for a charged BTZ black hole. There, it was implicitly assumed that the black hole mass was fixed when $\Lambda$ varies. However, the authors only fixed the parameters $m\,,q$ independently while the black hole mass is given by $M=\ft 14(m-\ft 14 q^2 \log L)$.  So their derivation of the thermodynamic volume is incorrect. Instead, we find $V_\pm=\pi r_\pm^2$ by using $V_{\mathrm{th}}=\big( \partial M/\partial P \big)_{S\,,Q}$ or our formal definition Eq.(\ref{first}).}

%In spite of that this cannot be proved in a general form, it is natural and reasonable. For example, considering a stationary AdS black hole, one finds for a fixed $\Lambda$
%\be \delta H_\infty=\delta M-\Omega_\alpha \delta J_\alpha\,,\qquad \delta H_+=T\delta S \,.\ee
%and
%\be dM=T dS+\Omega_\alpha dJ_\alpha \,.\ee
%The black hole mass turns out to be a function of the entropy and the angular momenta, namely $M=M(S\,,J_\alpha)$. However, when $\Lambda$ varies, the mass will also be a function of $P$, $M=M(S\,,J_\alpha\,,P)$ so that $V_{\mathrm{th}}=\big( \partial M/\partial P \big)_{S\,,J_\alpha}$. Thus, the $V_{\mathrm{th}}\delta P$ term in the Wald equation (\ref{waldgene}) can be understood as emerging from $\delta H_\infty^{reg}$. The divergent terms on both sides of the equation (\ref{waldgene}) should be exactly cancelled. Indeed, this was verified in [xxx] explicitly for several AdS black holes

\subsection{Complexity$=$action $2.0$}
To gain some physical insights from Eq.(\ref{volume0}), we first split the action into kinetic terms (including boundary terms) and non-derivative terms, namely
\bea
&&S_{tot}=S_{T}+S_{ND}\,,\quad \mathcal{L}_{tot}=T-U \,,\nn\\
&&S_T=\int_{\mathcal{M}}T+S_{GH}\,,\quad S_{ND}=-\int_{\mathcal{M}}U\,.
\eea
Our first assumption is the $\Lambda$ dependence of the action only appears in the non-derivative terms so
\be\label{thervolume} V=\fft 12 \int_{\mathcal{A}_{n-2}}\mathrm{d}x^1\mathrm{d}x^2\cdots \mathrm{d}x^{n-2}\int^{r_+}_{r_i}\mathrm{d}r\sqrt{-g}\,\fft{\partial U}{\partial \Lambda} \,.\ee
This is natural. The second assumption is the potential term $U$ depends linearly on $\Lambda$: $U=U_\Lambda+\cdots$, where $U_\Lambda\propto \Lambda$ (the coefficient in general is a function of the matter fields) and the dotted terms are independent of $\Lambda$. Hence, we have
\be\label{thervolume2} V= \int_{\mathcal{A}_{n-2}}\mathrm{d}x^1\mathrm{d}x^2\cdots \mathrm{d}x^{n-2}\int^{r_+}_{r_i}\mathrm{d}r\sqrt{-g}\,\fft{U_\Lambda}{ 2\Lambda} \,.\ee
Of course, this condition can be easily violated when the Lagrangian density contains additional coupling constants. For instance, if $\alpha$ is a coupling constant which has a length dimension $[L]^{-2\sigma}$, then one can redefine $\alpha=\bar \alpha\, \Lambda^\sigma$. In fact, this corresponds to choosing a different extended phase space so one naturally finds a different thermodynamic volume. However, as will be shown later, we are well motivated to study the thermodynamic volume defined in the extended phase space spanned by the linear $\Lambda$ term.

Under the above two assumptions, we find that the thermodynamic pressure multiplied by the thermodynamic volume is nothing else but a part of the non-derivative action (we call it $\Lambda$-action) evaluated on the WDW patch. We have
\be \dot S_{\Lambda_+} \equiv P V=-\fft{1}{16\pi G}\int_{\mathcal{A}_{n-2}}\mathrm{d}x^1\mathrm{d}x^2\cdots \mathrm{d}x^{n-2}\int^{r_+}_{r_i}\mathrm{d}r\sqrt{-g}\,U_{\Lambda} \,.\ee
For the black holes without an inner horizon, we have $r_i=0$. Otherwise, we can define $S_{\Lambda_-}$ for the inner horizon and find
\bea\label{ca2}
\dot S_{\Lambda}&\equiv& \dot S_{\Lambda_+}-\dot S_{\Lambda_-}=P\,(V_+-V_-)\nn\\
&=&-\fft{1}{16\pi G}\int_{\mathcal{A}_{n-2}}\mathrm{d}x^1\mathrm{d}x^2\cdots \mathrm{d}x^{n-2}\int^{r_+}_{r_-}\mathrm{d}r\sqrt{-g}\,U_{\Lambda}  \,,
\eea
which is still the same part of the growth of the action at late time evaluated on the WDW patch. Note that the $r_-\rightarrow 0$ limit reduces to the case without an inner horizon.

From these results, it becomes clear that the thermodynamic volume is in general not equal to the spatial volume of the WDW patch. The two conceptions just coincide by chance when the potential $U=\mathrm{cons}\times \Lambda+\cdots$. Indeed, the discussions in \cite{Couch:2016exn} were focused on black holes in Einstein-Maxwell theory (including pure Einstein gravity) which has $U=2\Lambda$. In this sense, it is not a surprise that in that paper the authors found the space-time volume of the WDW patch is related to the thermodynamic volume.

 Despite that the authors of \cite{Couch:2016exn} were not aware of the two conceptions are in general not equivalent, it has been shown in the paper that the ``complexity=volume 2.0" duality favors over the original CA-duality proposed by Brown et al in the sense that there are some cases for which the former respects the Lloyd bound whereas the latter does not.  However, now one may be confused by which of the two quantities: the spacetime volume or the thermodynamic volume, is a better notion for the gravity duals of quantum complexity. We would like to investigate this in details in the next section. Here we propose
\be\label{ca2duality}
\mathrm{complexity=action\,2.0}:\quad \mathcal{C}=\fft{S_{\Lambda}}{\pi \hbar} \,,
\ee
where $S_{\Lambda}$ is defined on the WDW patch and is not limited to late time. To avoid confusions, we refer it to ``complexity=action 2.0" (CA-2) duality. By contrast, the original ``complexity=volume 2.0" duality proposed in \cite{Couch:2016exn} is
\be\label{cv2duality}\mathrm{complexity=volume\,2.0}:\quad\mathcal{C}=\fft{1}{\pi \hbar}\,P\cdot \big(\mathrm{Spacetime \,\,Volume\big)}=\fft{S_{V}}{\pi \hbar}\,,\ee
where $S_V=S_{\Lambda}$ when $U_\Lambda=2\Lambda$. In particular, when $U_\Lambda\geq 2\Lambda$, we will have
\be S_V\leq S_\Lambda<S \,,\ee
where the second ``$<$" is guaranteed by the positivity of the kinetic action. So it is possible that for certain cases the CA duality does not respect the Lloyd bound whereas the CA-2 and CV-2 dualities does.

\section{Complexity$=$action $2.0$ vs complexity$=$volume $2.0$}\label{dualities}
\subsection{AdS planar black holes}
To investigate whether our new conjecture is reasonable, we first study AdS planar black holes. For Einstein-Scalar gravity, the scalar potential in general contains two classes of dimensional parameters: the cosmological constant $\Lambda$ and the other coupling constants $\Theta_\alpha$, both of which have length dimension $[L]^{-2}$. The first law of thermodynamics in the extended phase space reads
\be dM =T dS+V dP+\Upsilon_\alpha\,d\Theta_\alpha \,,\ee
where $\Upsilon_\alpha$ is the thermodynamic conjugate of $\Theta_\alpha$. Here for simplicity we have assumed the ``scalar charges" of the solutions do not contribute to the first law (This assumption is valid for all the hairy black hole solutions discussed in section \ref{application}. In fact, although it was established in \cite{Lu:2013ura} that the first law is significantly modified by the ``scalar charges", a recent paper \cite{Astefanesei:2018vga} shows that the first law remains unchanged if the black hole mass is properly redefined.  ). On one hand, by scaling dimensional arguments, one finds the Smarr relation is
\be\label{smarr1} (n-3)M=(n-2)T S-2V P-2\Upsilon_\alpha \Theta_\alpha \,,\ee
On the other hand, the AdS planar black holes have an extra scaling symmetry, which leads to a generalized Smarr relation \cite{Liu:2015tqa}
\be\label{smarr2} (n-1)M=(n-2)T S    \,.\ee
Combining the above results, we arrive at
\be M=P V+\Upsilon_\alpha \Theta_\alpha\geq PV \,,\ee
provided the thermodynamic conjugates $(\Upsilon_\alpha\,,\Theta_\alpha)$ are positive definite. Thus, in this case our CA-2 duality is always consistent with the Lloyd bound.

However, for charged planar black holes, the situation is not so simple. One has
\bea
&&(n-3)M=(n-2)T S+(n-3)\Phi_e Q_e-2V P-2\Upsilon_\alpha \Theta_\alpha \,,\nn\\
&&(n-1)M=(n-2)T S+(n-2)\Phi_e Q_e\,,
\eea
so that
\be M-\ft 12 \Phi_e Q_e= P V+\Upsilon_\alpha \Theta_\alpha \geq P V\,.\ee
However, the Lloyd bound is lowered as well by the charged carriers \cite{Brown:2015lvg} so the above equation does not lead to any definite conclusion. Indeed, in sec.\ref{chargedbh}, we will find that there exists certain charged black holes for which the CA-2 duality does not respect the Lloyd bound.

\subsection{Spherical black holes}
\begin{figure}[ht]
\centering
\includegraphics[width=230pt]{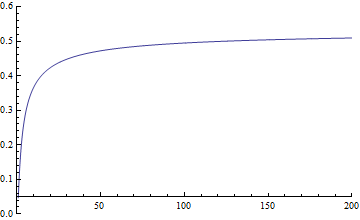}
\caption{{\it The ratio $\dot{S}_{\Lambda}/2M$ is plotted as a function of the black hole mass. We have set $g=1\,,\alpha=3\,,\omega_2=4\pi\,,G=1/4$.  }}
\label{zxffig0}\end{figure}

Spherical black holes do not have the generalized Smarr relation. We shall study them in a case-by-case basis.
First, for the black hole (\ref{zxf}), we obtain
\be \dot{S}_{\Lambda}=P V=\fft{\omega_2\,g^2}{16\pi G}r_+(r_++q)(2r_++q)\,,\quad  \dot{S}_{V}=\fft{\omega_2\,g^2}{16\pi G}r_+^2(2r_++3q)\,,\ee
which gives rise to
\be \dot{S}_{\Lambda}=\dot{S}_{V}+\fft{\omega_2\,g^2 q^2 r_+} {16\pi G}>\dot{S}_{V}\,,\ee
implying that the growth of the spatial volume of the WDW patch is smaller than $P V$. In Fig.\ref{zxffig0}, we plot $\dot S_{\Lambda}/2M$ as a function of the black hole mass. We find that $\dot S_{\Lambda}=PV<2M$ throughout the mass range. Thus, both the CA-2 and CV-2 dualities are consistent with the Lloyd bound.

For the black hole (\ref{zxf2}), we obtain
\bea
&&\dot{S}_{\Lambda}=\fft{\omega_2\,g^2}{16\pi G}r_+\Big( 2r_+^2+3(1-\mu)q r_++(2\mu^2-3\mu+1)q^2 \Big)\,,\nn\\
&&\dot{S}_{V}=\fft{\omega_2\,g^2}{16\pi G}\Big[ (r_++q)\Big(2r_+^2+(5-\mu)q r_++(\mu^2-2\mu+3)q^2\Big)\nn\\
&&\qquad \qquad \qquad\qquad\qquad\quad+\ft{\mu(1-\mu^2)}{\mu+2}q^3\,{}_2F_1\big[1\,,\mu+2\,,\mu+3\,,\ft{r_+}{r_++q} \big]   \Big]\,.
\eea
These expressions do not have simple relations to the black hole mass, which is given by
\be M=\fft{\mu\omega_2}{24\pi G}\,q\Big((g^2-\alpha^2)(1-4\mu^2)q^2-3k \Big) \,.\ee
\begin{figure}[ht]
\centering
\includegraphics[width=230pt]{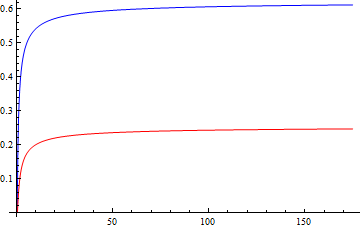}
\caption{{\it The ratios $\dot{S}_{\Lambda}/2M$ (Blue) and $\dot{S}_{V}/2M$ (Red) are plotted as functions of the black hole mass. We have set $\alpha=0\,,\mu=1/4\,,\omega_2=4\pi\,,G=1\,,g=1$.  }}
\label{zxffig}\end{figure}
To simplify the discussions, we focus on $\alpha=0$ and $\mu<1/2$. In this case, the horizon radii can be solved analytically as
\be r_+=\fft 12\Big(\,\sqrt{(1-4\mu^2)q^2-4k\ell^2} -(1-2\mu)q\Big) \,.\ee
Note that for a spherical black hole, the positivity of the mass $M$ and $r_+$ provides a lower bound on the scalar charge: $q\geq q_c$ with $q_c=\ell/\sqrt{\mu(1-2\mu)}$. With these results in hand, we are ready to plot $\dot{S}_{\Lambda}$ and $\dot{S}_{V}$ as a function of the black hole mass.
This is shown in Fig.\ref{zxffig}. We again find
\be \dot{S}_{V}<\dot{S}_{\Lambda}<2M \,,\ee
throughout the parameters space.

\subsection{Charged black holes}\label{chargedbh}
The presence of conserved charges puts additional constraints on the system so the growth of the complexity will be slower than the case without the charges. Hence, the Lloyd bound should be properly generalized. A natural choice is \cite{Brown:2015lvg}
\be\label{chargedbound} \mathcal{\dot C}\leq \fft{2}{\pi \hbar}\Big( \big(M-\mu_e Q\big)-\big(M-\mu_e Q\big)_{gs} \Big) \,,\ee
where $\mu_e$ denotes the chemical potential. The term $\big(M-\mu_e Q\big)_{gs}$ is simply $\big(M-\mu_e Q\big)$ evaluated for the ground state. In this subsection, we would like to study the complexity growth of charged AdS black holes. The ground state is taken as either an empty AdS space (for grand canonical ensemble) or an extremal black hole (for canonical ensemble). We will work in the grand canonical ensemble because the charged black holes studied in this subsection do not have an inner horizon.

We first consider the four dimensional Kaluza-Klein theory
\be \mathcal{L}=R-\ft 12\big(\partial \phi \big)^2-\ft 14 e^{\sqrt{3}\phi}F^2+6g^2\cosh{\Big(\ft{\phi}{\sqrt{3}}\Big)} \,.\ee
The electrically charged black hole solution is given by
\bea\label{kksol}
&&ds^2=-H^{-1/2}f dt^2+H^{1/2}\Big(dr^2/f+r^2 d\Omega_{2\,,k}^2 \Big)\,,\quad \phi=\ft{\sqrt{3}}{2}\log{H}\,,\nn\\
&&H=1+\fft{2\mu q}{r}\,,\quad f=g^2 r^2 H+k-\fft{2\mu}{r}\,,\quad A=2\mu\Big(\ft{\sqrt{q(1+k q)}}{r_++2\mu q}- \ft{\sqrt{q(1+k q)}}{r+2\mu q}\Big)dt\,,
\eea
where $\mu\,,q$ are two integration constants. The black hole mass and the electric charge are given by
\be M=\fft{\omega_2}{8\pi G}\,\mu (2+k q)\,,\qquad Q=\fft{\omega_2}{8\pi G}\,\mu \sqrt{q\big(1+k q\big)} \,.\ee
Note that for a spherical black hole, the positivity of $\mu$ and $q$ leads to an upper bound on the electric charge $Q<M$.

First, we calculate the action growth by using Eq.(\ref{sdot}). We find
\be \dot S=\fft{\omega_2\,\mu\Big((k q+4)r_++6\mu q\Big)}{8\pi G\big(r_++2\mu q\big)} \,,\ee
and
\be \dot S-2\big(M-\mu_e Q\big)=\fft{\omega_2\mu q(2\mu-k r_+)}{8\pi G(r_++2\mu q)}>0\,,\ee
which is positive definite due to the horizon condition $f(r_+)=0>k-2\mu/r_+$. Thus, in this case the CA duality always violates the Lloyd bound.

\begin{figure}[ht]
\centering
\includegraphics[width=230pt]{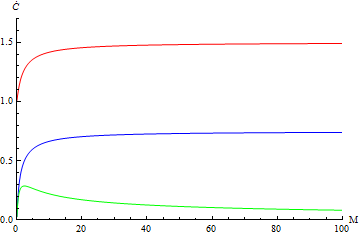}
\caption{{\it The growth of the complexity normalized by the Lloyd bound is plotted as a function of the black hole mass for the Kaluza-Klein black hole. The red/blue/green lines correspond to CA/CA-2/CV-2 dualities respectively. We have set $k=1\,,\omega=4\pi\,,G=1\,,g=1\,,q=1$.  }}
\label{kk4fig}\end{figure}

On the contrary, for the CA-2 and CV-2 dualities, we find
\be \dot{S}_{V}<\dot{S}_{\Lambda}=\fft{\omega_2(2r_++3\mu q)(2\mu-k r_+)}{16\pi G(r_++2\mu q)} \,,\ee
and
\bea
\dot{S}_{\Lambda}-2\big(M-\mu_e Q \big)&=&-\fft{\omega_2\Big( 2k r_+^2+(7k q+4)\mu q+2\mu^2 q \Big)}{16\pi G(r_++2\mu q)}<0 \,,
\eea
 for both spherical and planar black holes. To help gain intuition, in Fig.\ref{kk4fig}, we plot the growth of the complexity as a function of the black hole mass for the three dualities.

Next, we consider the singly charged black hole in the $\mathcal{N}=4\,,D=4$ gauged supergravity
\be \mathcal{L}=R-\ft 12\big(\partial \phi \big)^2-\ft 14 e^{\phi}F^2+2g^2\big(2+\cosh{\phi} \big) \,.\ee
The black hole reads
\bea\label{n4d4}
&&ds^2=-f dt^2+dr^2/f+r(r+q) d\Omega_{2\,,k}^2 \,,\quad \phi=\log{\Big( 1+\fft{q}{r} \Big)}\,,\nn\\
&&f=g^2 r(r+q) +k-\ft{8\pi G M}{\omega_2(r+q)}\,,\quad A=\ft{16\pi G}{\omega_2}\Big( \ft{Q}{r_++q}-\ft{Q}{r+q} \Big)dt\,.
\eea
Here $M$ and $Q$ are the black hole mass and the electric charge, which are related by
\be M=\fft{16\pi G}{\omega_2 q}\,Q^2 \,.\ee
By simple calculations, we deduce
\bea
&&\dot S=\fft{2r_++q}{r_++q}M-\fft{k\omega_2 q}{8\pi G}\,,\nn\\
&&\dot S_{\Lambda}=\fft{2r_++q}{2(r_++q)}M-\fft{k\omega_2(2r_++q)}{16\pi G}\,,\nn\\
&&\dot S_{V}=\fft{r_+(2r_++3q)}{2(r_++q)^2}M-\fft{k\omega_2 r_+(2r_++3q)}{16\pi G(r_++q)}\,.
\eea
For the CA and CV-2 dualities, we find
\bea
&&\label{ndca}\dot S-2\big(M-\mu_e Q \big)=\fft{q}{r_++q}M-\fft{k\omega_2 q}{8\pi G}>0\,,\\\nn
&&\dot S_{V}-2\big(M-\mu_e Q \big)=-\fft{r_+(2r_++q)}{2(r_++q)^2}M-\fft{k\omega_2 r_+(2r_++3q)}{16\pi G(r_++q)}<0  \,,
\eea
for spherical/planar black holes. Here the sign of the first equation is guaranteed by the existence of the event horizon which requires $k-\ft{8\pi G M}{\omega(r_++q)}<0$.
However, the situation of the CA-2 duality is slightly involved. We find
\be
\label{ndca2}\dot{S}_{\Lambda}-2\big(M-\mu_e Q \big)=-\fft{2r_+-q}{2(r_++q)}M-\fft{k\omega_2(2r_++q)}{16\pi G}\,.
\ee
\begin{figure}[ht]
\includegraphics[width=140pt]{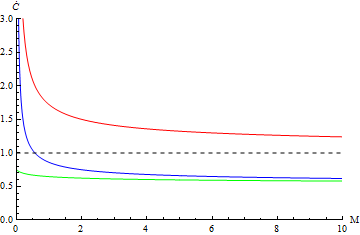}
\includegraphics[width=140pt]{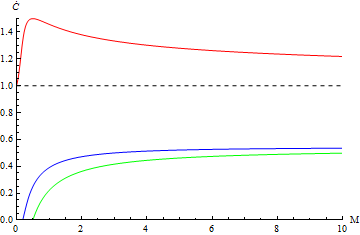}
\includegraphics[width=140pt]{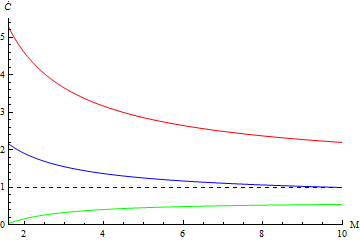}
\caption{{\it The growth of the complexity normalized by the Lloyd bound for the planar (left) and spherical black hole (middle and right) is plotted as a function of the black hole mass. The red/blue/green lines correspond to CA/CA-2/CV-2 dualities respectively. We have set $q=1$ for the left and middle panels and $q=3$ for the right panel. Some other parameters has been chosen as $\omega_2=4\pi\,,G=1\,,g=1$.}}
\label{n4d4fig}\end{figure}
 Its sign strongly depends on the parameters of the solutions. For the planar black hole, Eq.(\ref{ndca2}) will be positive definite for a small black hole with $r_+<q/2$, as shown in the left panel of Fig. \ref{n4d4fig}. For the spherical black hole, we find $\dot S_{\Lambda}$ will exceed the Lloyd bound as well for a relatively large $q$, as shown in the right panel of Fig.\ref{n4d4fig}.

From these results, we may conclude that the CV-2 duality is slightly better than the CA-2 duality. In fact, its various advantages suggest that it might be the best holographic dual of the quantum complexity of the boundary field theories.

\section{Conclusion}

In this paper, we investigate various proposals (the CA, CA-2 and CV-2 dualities) for the gravity duals of quantum complexity of the boundary field theories.

For the original CA duality, we calculate the action growth of the WDW patch for generally static black hole solutions in Einstein gravity by adopting the rigorous method developed in \cite{Lehner:2016vdi}. We prove that the growth of the action of the Wheeler-DeWitt (WDW) patch at late time is essentially given by the bulk action evaluated at the black hole interior together with the Gibbons-Hawking boundary terms evaluated on the (inner and outer) event horizon. The proof shows some universal features which do not rely on the details of the gravity theories. We thus argue that the above argument is valid for general higher derivative gravities, without knowing the explicit form of the joint actions.

By carefully examining the thermodynamic volume using Wald-Iyer formalism, we relate its product with the thermodynamic pressure (PV) to a part of the non-derivative gravitational action evaluated on the WDW patch, which is linearly proportional to the cosmological constant. Considering the physical relevance of the thermodynamic volumes of AdS black holes, we propose a new gravity dual for the complexity referred to ``complexity=action 2.0" (CA-2) duality: $\mathcal{C}= S_{\Lambda}/\pi\hbar$, where $S_\Lambda$ reduces to $P V$ in the stationary limit.

We then calculate the growth of the complexity at late time by using the various dualities for a number of black holes in Einstein-Maxwell-Dilaton theories. We find that in many cases, the CA-2 and CV-2 dualities respect the Lloyd bound whereas the CA duality does not. In this sense, the CA-2 and CV-2 dualities favor over the CA-duality. However, we also find a counter example: the charged black hole in $\mathcal{N}=4\,,D=4$ supergravity, for which the CA-2 duality violates the Lloyd bound as well. We thus argue that the CV-2 duality might be the best holographic dual of the complexity of the boundary field theory.

There are many new interesting directions that are deserved further investigations. For example, modifying the Lloyd bound by an overall factor, studying the complexity formation or exploring the time evolution of the complexity in dynamical space-times, to see what happens in the new situations for the above three dualities.

Certainly, in order to find the correct holographic dual of the complexity, it is of great importance to properly define and calculate the complexity from the field theory perspective. We refer the readers to \cite{Jefferson:2017sdb,Chapman:2017rqy,Yang:2017nfn,Khan:2018rzm,Hackl:2018ptj,Yang:2018nda,chapman:tfd,minyong:coherent} for recent developments in this direction.

\section*{Acknowledgments}
We would like to thank Shira Chapman, Hugo Marrochio, Juan Pablo Hernandez,  Robert C. Myers, Shan-Ming Ruan and Yijian Zou for useful comments and discussions.
Z.Y. Fan is supported in part by the National Natural Science Foundations of China (NNSFC) with Grant No. 11575270 and also supported by Guangdong Innovation Team for Astrophysics(2014KCXTD014).
M. Guo is supported in part by NNSFC Grants No. 11775022 and No. 11375026 and also supported by the China Scholarship Council. M. Guo also thanks the Perimeter Institute \lq\lq{}Visiting Graduate Fellows\rq\rq{} program. Research at Perimeter Institute is supported by the Government of Canada through the Department of Innovation, Science and Economic Development and by the Province of Ontario through the Ministry of Research, Innovation and Science.\\

\end{document}